\documentclass[aps,prl,twocolumn,superscriptaddress]{revtex4}

\usepackage{graphicx}
\usepackage{amsmath}
\usepackage{amssymb}

\begin{document}

\title{Orbital Feshbach Resonance: A ``Wide" Narrow Resonance for Higher Transition Temperature Fermi Superfluid}
\author{Junjun Xu}
\thanks{They contribute equally to this work}
\affiliation{Department of Physics, University of Science and Technology Beijing, Beijing 100083, China}
\author{Ren Zhang}
\thanks{They contribute equally to this work}
\affiliation{Institute for Advanced Study, Tsinghua University, Beijing, 100084, China}
\author{Yanting Cheng}
\affiliation{Institute for Advanced Study, Tsinghua University, Beijing, 100084, China}
\author{Peng Zhang}
\email{pengzhang@ruc.edu.cn}
\affiliation{Department of Physics, Renmin University of China, Beijing, 100872,
China}
\affiliation{Beijing Key Laboratory of Opto-electronic Functional Materials \&
Micro-nano Devices, 100872 (Renmin Univeristy of China)}
\affiliation{Beijing Computational Science Research Center, Beijing, 100084, China}
\author{Ran Qi}
\email{qiran@ruc.edu.cn}
\affiliation{Department of Physics, Renmin University of China, Beijing, 100872,
China}
\author{Hui Zhai}
\email{hzhai@tsinghua.edu.cn}
\affiliation{Institute for Advanced Study, Tsinghua University, Beijing, 100084, China}

\date{\today }

\begin{abstract}

In this letter we show that the recently theoretically predicted and experimentally observed ``orbital Feshbach resonance" in alkali-earth-like ${}^{173}$Yb atom is a narrow resonance in energy, while it is hundreds Gauss wide in term of magnetic field strength, taking the advantage that the magnetic moment difference between the open and closed channels is quite small.  Therefore this is an ideal platform for the experimental realization of a strongly interacting Fermi superfluid with narrow resonance. We show that the transition temperature for the Fermi superfluid in this system, especially at the BCS side of the resonance, is even higher than that in a wide resonance, which is also due to the narrow character of this resonance. Our results will encourage experimental efforts to realize Fermi superfluid in the alkali-earth-like ${}^{173}$Yb system, the properties of which will be complementary to extensively studied Fermi superfluids nearby a wide resonance in alkali ${}^{40}$K and ${}^6$Li systems.

\end{abstract}

\maketitle

Pairing of fermions is a universal mechanism for both superconductivity in materials and Fermi superfluid in neutral atoms. In the weakly attractive BCS regime, the transition temperature $T_\text{c}$ is much smaller than the Fermi temperature $T_\text{F}$ \cite{BCS}. In cold atom systems, the Feshbach resonance (FR) provides a tool to significantly enhance the attraction between atoms \cite{Feshbach,Feshbach2}, with which the $T_\text{c}$ can be increased to the order of $0.1T_\text{F}$ \cite{Unitary,review2}. So far, in term of $T_\text{c}/T_\text{F}$, this is the highest transition temperature ever achieved and in the past decade or so it has been realized and extensively studied by many laboratories with ${}^{40}$K and ${}^{6}$Li atoms \cite{Unitary,review2}.

Interactions between ultracold atomic gases are usually dominated by $s$-wave scattering and can be well described by the $s$-wave phase shift $\theta_k$, which 
can be expanded as  \cite{Feshbach,Feshbach2,Unitary}
\begin{align}
k\cot \theta_k \simeq-\frac{1}{a_s}+\frac{1}{2}r_0 k^2. \label{expansion}
\end{align}
where $k$ is the relative momentum between two atoms. 
The leading order term gives the scattering length $a_\text{s}$ which diverges at a resonance. The next order term is characterized by an effective range $r_0$, which describes how fast the scattering phase shift changes in energy. Effective range also controls how sensitive the $a_s$ depends on the magnetic field strength. Normally around a FR the magnetic field dependence of $a_s$ can be casted into
\begin{align}
a_{s}=a_\text{bg}\left(1-\frac{\Delta_\text{B}}{B-B_\text{res}}\right), \label{a_B}
\end{align}
where $B_\text{res}$ is the field location of a FR, $a_\text{bg}$ is so-called the background scattering length, $\Delta_\text{B}$ is the resonance width. Neglecting the short range van de Waals physics, $\Delta_\text{B}$ is related to $r_0$ via \cite{width1,width2}
\begin{align}
r_0= -\frac{2\hbar^2}{m\delta\mu\Delta_\text{B} a_\text{bg}},  \label{r0_DB}
\end{align}
where $m$ is the single-atom mass and $\delta\mu$ is the magnetic moment difference between the open and closed channels.

A resonance can be classified as a wide or narrow resonance, depending on how $r_0$ compares with the characteristic length scales of the system. This length scale could be either van der Waals length $r_\text{vdW}$ or the inverse of the Fermi momentum $1/k_\text{F}$. One can introduce a parameter called $s_\text{res}=8\pi r_\text{vdW}/(\Gamma(1/4)^2|r_0|)$, and resonances with $s_{\rm res}\ll 1$ are called narrow resonances \cite{Feshbach2}. 
So far all experiments on fermion superfluid nearby a FR are performed with a wide resonance, such as ${}^{6}$Li (at 832G) with $s_\text{res}=59 $  and ${}^{40}$K (at 224.2G) with $s_\text{res}=2.7$ \cite{Feshbach2}. According to Eq. (\ref{r0_DB}), the larger $|r_0|$ the smaller $\Delta_{\rm B}$. Therefore,
one major difficulty for experiments with a narrow magnetic FR of alkali atoms is that the small $\Delta_\text{B}$ requires precise controllability and ultra-stability of the magnetic field in order to stay on strongly interacting regime. A number of experiments have studied strongly interacting Fermi gases nearby a narrow FR \cite{Hulet, Grimm,Hara}, and despite of the novel effects revealed by these experiments, the fermion superfluid has not been achieved there.

Recently, a new kind of FR -- orbital Feshbach resonance (OFR) -- has been predicted for scattering between the electronic ground state ${}^1S_0$ and the excited clock state ${}^{3}P_0$ of ${}^{173}$Yb atoms \cite{Zhang}, and it has been experimentally observed from measurements including the thermalization rate, the atom loss rate and the anisotropic expansion \cite{Munich, Florence}. Moreover, the observed lifetime of atomic gases at OFR is rather long \cite{Munich, Florence}, which makes strongly interacting Fermi superfluid at OFR attainable. To motivate experiments to pursue this direction, two questions have to be answered. First, whether $T_\text{c}$ at OFR is also as high as $T_\text{c}$ of the Fermi superfluid studied before? Second, whether this new superfluid in ${}^{173}$Yb has different properties comparing to that studied in alkali atoms of ${}^{40}$K and ${}^6$Li before? The goal of this letter is to provide positive answers to these questions, and the answers are closely related to the effective range and the width of OFR.

\textit{Effective Range of OFR.} Let us first briefly review the two-body scattering property for OFR \cite{Zhang}. Considering two atoms, one is in ${}^1S_0$ state denoted by $|g\rangle$ and the other in the ${}^3P_0$ state denoted by $|e\rangle$, and they are in different nuclear spin state denoted by $|\uparrow\rangle$ and $|\downarrow\rangle$ respectively. These two atoms can either stay in the open channel, denoted by $|o\rangle=(|g\downarrow;e\uparrow\rangle-|e\uparrow;g\downarrow\rangle)/\sqrt{2}$, or in the closed channel, denoted by $|c\rangle=(|g\uparrow;e\downarrow\rangle-|e\downarrow;g\uparrow\rangle)/\sqrt{2}$. The energy of the open and the closed channels are offset by $\delta=\delta\mu B$, where is due to the difference of the nuclear Land\'e $g$ factor between ${}^1S_0$ and  ${}^3P_0$ states and $\delta \mu$ is as small as $2\pi\hbar\times 112 \text{Hz}/(\text{G}\times \Delta_m)$ for ${}^{173}$Yb, and $\Delta_m$ is the difference in nuclear spin quantum number. The free Hamiltonian is given by
\begin{equation}
\hat{H}_0=\left(-\frac{\hbar^2\nabla^2}{m}+\delta\right)|c\rangle\langle c|-\frac{\hbar^2\nabla^2}{m}|o\rangle\langle o|.
\end{equation}
On the other hand, the interaction Hamiltonian is diagonalized in the $|\pm\rangle$ bases which are defined as
\begin{equation}
|\pm \rangle=\frac{1}{2}(|ge\rangle\pm |eg\rangle)(|\uparrow\downarrow\rangle\mp|\downarrow\uparrow\rangle)=\frac{1}{\sqrt{2}}(|c\rangle\mp |o\rangle),
\end{equation}
and the interaction Hamiltonian contains two independent scattering lengths $a_{s+}$ and $a_{s-}$ as
\begin{equation}
\hat{V}=\left(\frac{4\pi\hbar^2}{m}\sum_{i=\pm}a_{si}|i\rangle\langle i|\right)\delta({\bf r})\frac{\partial}{\partial r}(r\cdot),\label{v}
\end{equation}
where ${\bf r}$ is the relative position between atoms. 
The validity of pseudo-potential description in this system has also been carefully examined in Ref. \cite{Cheng}.

\begin{figure}[tp]
\includegraphics[width=3.2 in]
{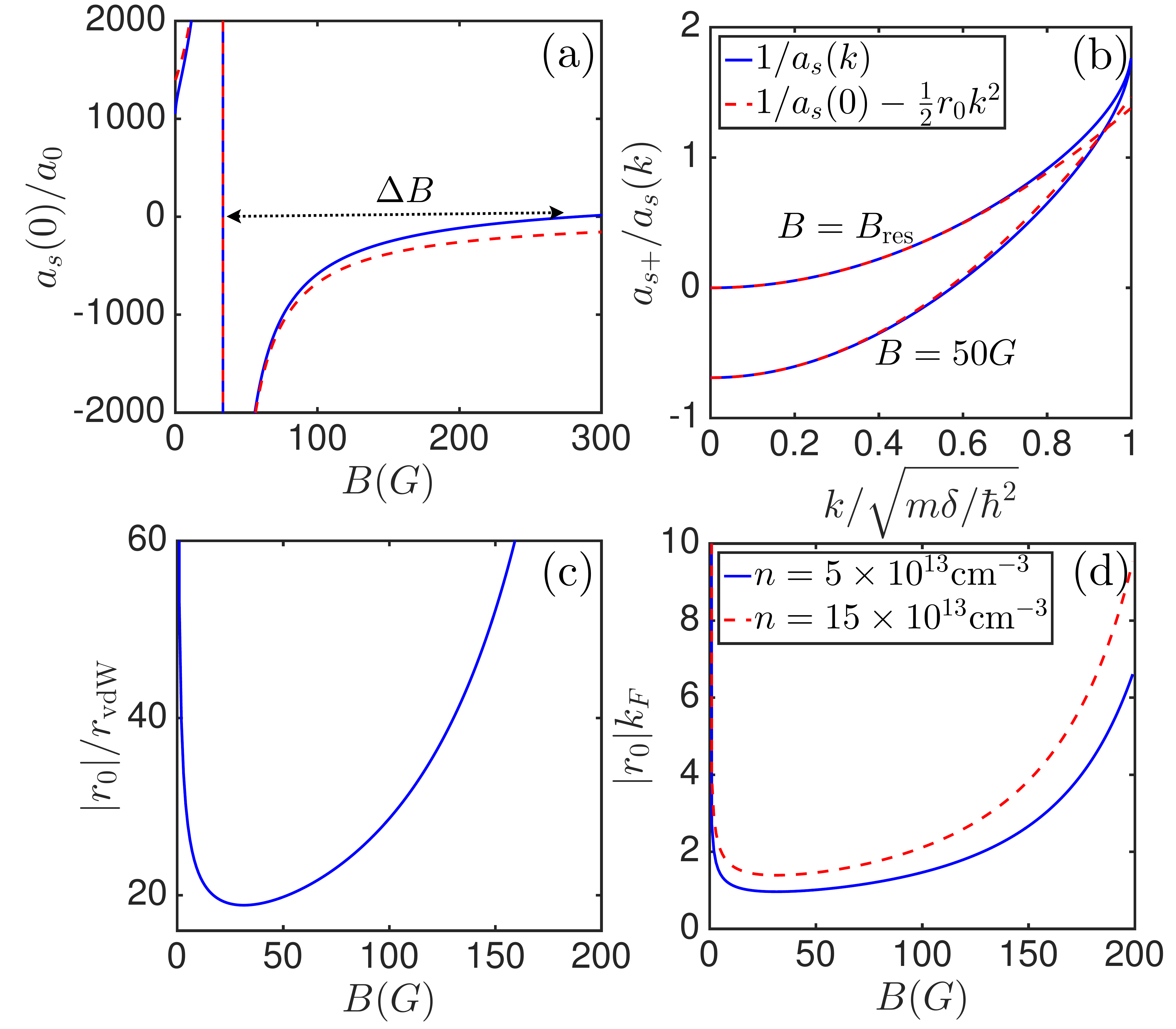}
\caption{(a): The scattering length $a_s$ (in unit of Bohr radius $a_0$), the solid line is the rigorous result Eq. (\ref{aa}) and the dashed line is the expansion with formula Eq. (\ref{a_B}); (b) $a_{s+}/a_s(k)$ as a function of  $k$ for $B=B_\text{res}=33.5{\rm G}$ and $B=50$G (BCS side); (c) the absolute value of effective range $|r_0|/r_\text{vdW}$; and (d) $k_\text{F}|r_0|$, the solid and the dash lines correspond to two different densities. (a), (c) and (d) are functions of  the magnetic field strength $B$. Here we have taken $a_{s+}=1900a_0$ and $a_{s-}=200a_0$ as reported in Ref.\cite{Munich,Florence}.  $r_\text{vdW}=84.8a_0$ \cite{rvdw}, $\Delta_m=5$. }
\label{range}
\end{figure}

The fact that the single-particle term is diagonalized in $|c\rangle$ and $|o\rangle$ bases while the interaction term is diagonalized in the $|\pm\rangle$ bases provides coupling between two channels, and the largeness of at least one of $a_{s\pm}$ provides a shallow bound state accessible by realistic magnetic field strength even with such a small $\delta\mu$. These are two essential ingredients for this OFR to be experimentally achievable. With this model, one can obtain the scattering length \cite{Zhang}
\begin{equation}
a_{s}=\frac{-a_{s0}+\sqrt{m\delta/\hbar^2}(a^2_{s0}-a^2_{s1})}{a_{s0}\sqrt{m\delta/\hbar^2}-1},\label{aa}
\end{equation}
where $a_{s0}=(a_{s+}+a_{s-})/2$ and $a_{s1}=(a_{s-}-a_{s+})/2$. $a_s$ as a function of the magnetic field strength is shown in Fig. \ref{range}(a). The divergent $a_\text{s}$ at $\delta_\text{res}=\hbar^2/(ma^2_{s0})$ gives rise to OFR. Below, we will show that, although the OFR is not an extremely narrow case, it has a flavor of narrow resonance for a typical density of cold atomic gases. 

Firstly, we can show that, by expanding $\delta$ around $\delta_\text{res}$, and after some straightforward calculations \cite{supple}, Eq. (\ref{aa}) can also be casted into the form as Eq. (\ref{a_B}), where
\begin{align}
&a_\text{bg}=\frac{2a^2_{s0}-3a^2_{s1}}{2a_{s0}},\label{abg}\\
&B_\text{res}=\frac{\hbar^2}{\delta\mu m a^2_{s0}},\label{Bres}\\
&\Delta_{\rm B}=\frac{4\hbar^{2}a^2_{s1}}{\delta\mu ma^2_{s0}(2a^2_{s0}-3a^2_{s1})}.\label{DeltaB}
\end{align}
The expansion is shown as the dashed line of Fig. \ref{range}(a).
With recently updated scattering lengths for ${}^{173}$Yb \cite{Munich,Florence}, taking $a_{s+}=1900a_0$ and $a_{s-}=200a_0$ and for $\Delta_m=5$, we obtained $B_\text{res}=33.5$G, $a_\text{bg}=18a_0$ and $\Delta_{\rm B}=2585$G. Note that with Eq. (\ref{aa}), if one defines $\Delta_{\rm B}$ as the difference between the zero-crossing and the resonance point, one has $\Delta_{\rm B}=249$G. The difference between these two values of $\Delta_{\rm B}$ is due to the fact that the expansion Eq. (\ref{a_B}) is invalid near the zero-crossing. 

Secondly, with the pseudo-potential model it is also straightforward to calculate the $\theta_k$ for finite $k$. One can define an energy dependent scattering length via $-1/a_{s}(k)\equiv k\cot\theta_k$. In Fig. \ref{range}(b) we plot $1/a_{s}(k)$ as a function of $k$ for two typical magnetic field strengths, and one can see significant variation of $a_s(k)$ over a typical range of the Fermi energy. By performing a low-energy expansion as Eq. (\ref{expansion}), we obtain a negative $r_0$ as \cite{supple}
\begin{equation}
r_0=-\frac{a_{s1}^{2}}{\sqrt{m\delta/\hbar^{2}}\left[a_{s0}-\sqrt{m\delta/\hbar^{2}}\left(a_{s0}^{2}-a_{s1}^{2}\right)\right]^{2}}.
\end{equation}
One can verify that $r_0$ at $\delta=\delta_\text{res}$, as well as $\Delta_{\rm B}$ and $a_\text{bg}$ introduced in Eqs. (\ref{abg}, \ref{DeltaB}) satisfy the relation Eq. (\ref{r0_DB}). Using the van der Waals length of ${}^{173}$Yb, we plot $|r_0|/r_\text{vdW}$ as a function of the magnetic field strength in Fig. \ref{range}(c), and typically it is much larger than 1. At resonance, one can obtain that $s_{\text{res}}=0.1$. In Fig. \ref{range}(d), we plot $k_\text{F}|r_0|$ as a function of the magnetic field strength and it is $\sim 1$ nearby OFR.  In addition, as another manifestation of narrow resonance feature, the binding energy of the Feshbach molecule does not follow $-\hbar^2/(ma^2_s)$ once slightly away from resonance \cite{supple}.

All discussion above clearly show that the OFR does not belong to a wide resonance. Concequently, a single $k$-independent scattering length is not enough to quantitatively capture this system, for instance, as we shall see in the following discussion of $T_\text{c}$. On the other hand, in term of the magnetic field, the width $\Delta_{\rm B}$ of the OFR is of a few hundred Gauss, which takes the advantage of very small $\delta\mu$ in this system. This makes the experimental study much easier.

\textit{Transition Temperature of Fermi Superfluid nearby OFR.} Now we consider the many-body Hamiltonian for a Fermi gas nearby OFR, which can be written as \cite{Zhang}
\begin{align}
&\hat{H}=\hat{H}_{0\text{o}}+\hat{H}_{0\text{c}}+\sum_{\mathbf{q}}\left[\frac{g_+}{2}\hat{A}^\dag_{+,\mathbf{q}}\hat{A}_{+,\mathbf{q}}+\frac{g_-}{2}\hat{A}^\dag_{-,\mathbf{q}}\hat{A}_{-,\mathbf{q}}\right],\\
&\hat{H}_{0\text{o}}=\sum_{{\bf k}}\varepsilon^\text{o}_{\bf k}(c_{g\downarrow {\bf k}}^{\dagger}c_{g\downarrow {\bf k}}+c_{e\uparrow {\bf k}}^{\dagger}c_{e\uparrow {\bf k}})\\
&\hat{H}_{0\text{c}}=\sum_{{\bf k}}\varepsilon^\text{c}_{\bf k}(c_{g\uparrow {\bf k}}^{\dagger}c_{g\uparrow {\bf k}}+c_{e\downarrow {\bf k}}^{\dagger}c_{e\downarrow {\bf k}}),
\end{align}
where $\varepsilon^\text{o}_{\bf k}=\hbar^2{\bf k}^2/(2m)-\mu$, $\varepsilon^\text{c}_{\bf k}= \hbar^2{\bf k}^2/(2m)+\delta/2-\mu$ and
\begin{align}
&\hat{A}_{+,\mathbf{q}}=\sum_{{\bf k}}(c_{g\uparrow{\bf q}/2-{\bf k}}c_{e\downarrow{\bf q}/2+{\bf k}}\!-\!c_{g\downarrow{\bf q}/2-{\bf k}}c_{e\uparrow{\bf q}/2+{\bf k}})\\
&\hat{A}_{-,\mathbf{q}}=\sum_{{\bf k}}(c_{g\uparrow{\bf q}/2-{\bf k}}c_{e\downarrow{\bf q}/2+{\bf k}}\!+\!c_{g\downarrow{\bf q}/2-{\bf k}}c_{e\uparrow{\bf q}/2+{\bf k}}),
\end{align}
with renormalization relations $1/g_{\pm}=m/(4\pi\hbar^2 a_{s\pm})-\sum_{k}m/(\hbar^2 {\bf k}^2)$. Written into the open and the closed channel bases, the interaction term contains the intra-channel and inter-channel scatterings with coupling constants $g_0=(g_{+}+g_{-})/2$ and $g_{1}=(g_{-}-g_{+})/2$, respectively.

\begin{figure}[tp]
\includegraphics[width=3.0 in]
{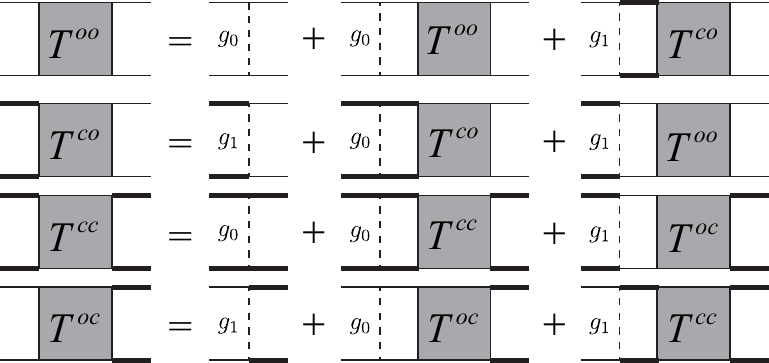}
\caption{Diagram for many-body T-matrix for a Fermi gas at OFR. The solid and thick solid lines denote the single particle propagators of atom in the open and closed channels respectively.}
\label{diagram}
\end{figure}

As emphasized in Ref. \cite{Zhang}, in OFR, the energy offset between the open and the closed channel is comparable to the Fermi energy, and therefore the scattering states in the closed channel $|c\rangle$ should be also considered. Hence, the theory for OFR is also different from previous studies of two-channel model for narrow FRs, where only a point-particle bound state is included in the closed channel, and is referred as atom-molecule two-channel model \cite{He}. Therefore, in our system, the many-body $T$-matrix should contain $T^{\text{oo}}$, $T^{\text{oc}}$, $T^{\text{co}}$ and $T^{\text{cc}}$, with the incoming and the outgoing states being in either the open channel or the closed channel, respectively. Following the diagrams shown in Fig. \ref{diagram}, they satisfy a set of the Dyson equations
\begin{align}
&T^{\text{oo}}=g_0+g_0\chi^\text{o}T^{\text{oo}}+g_1\chi^\text{c}T^{\text{co}}, \label{OO}\\
&T^{\text{co}}=g_1+g_0\chi^\text{c}T^{\text{co}}+g_1\chi^\text{o}T^{\text{oo}}, \label{CO}\\
&T^{\text{cc}}=g_0+g_0\chi^\text{c}T^{\text{cc}}+g_1\chi^\text{o}T^{\text{oc}}, \label{CC}\\
&T^{\text{oc}}=g_1+g_0\chi^\text{o}T^{\text{oc}}+g_1\chi^\text{c}T^{\text{cc}}, \label{OC}
\end{align}
where
\begin{align}
\chi^{\text{o/c}}({\bf q},\omega)=\sum_{\bf k}\frac{1-f^{\text{o/c}}_{{\bf q}/2+{\bf k}}-f^{\text{o/c}}_{{\bf q}/2-{\bf k}}}{\omega+i0^+-\varepsilon_{{\bf q}/2+{\bf k}}^{\text{o/c}}-\varepsilon_{{\bf q}/2-{\bf k}}^{\text{o/c}}},
\end{align}
and $f^{\text{o/c}}_{\mathbf{k}}=[\exp(\varepsilon^{\text{o/c}}_{\bf k})+1]^{-1}$ is the Fermi distribution functions. The solution of Eq. (\ref{OO}-\ref{OC}) gives
\begin{align}
T^{\text{oo}}&=\frac{g_0-\left(g_0^2-g_1^2\right)\chi^\text{c}}{1-g_0(\chi^\text{o}+\chi ^\text{c})+(g_0^2-g_1^2)\chi^o\chi^c},\\
T^{\text{cc}}&=\frac{g_0-\left(g_0^2-g_1^2\right)\chi^\text{o}}{1-g_0(\chi^o+\chi ^\text{c})+(g_0^2-g_1^2)\chi^\text{o}\chi^\text{c}},\\
T^{\text{oc}}&=T^{\text{co}}=\frac{g_1}{1-g_0(\chi^\text{o}+\chi ^\text{c})+(g_0^2-g_1^2)\chi^\text{o}\chi^\text{c}}.
\end{align}
Therefore, the pole of the T-matrix at the zero momentum and the zero frequency gives the Thouless criterion for this system
\begin{equation}
1-g_0\left(\chi^{\text{o}}_0+\chi^{\text{c}}_0\right)+\left(g_0^2-g_1^2\right)\chi^{\text{o}}_0\chi^{\text{c}}_0=0, \label{Thouless}
\end{equation}
where $\chi^{\text{o/c}}_0\equiv\chi^{\text{o/c}}(0,0)$.
In addition, we also need a number equation to determine the change of chemical potential in the normal state due to the interaction effect. Following the spirt of NSR \cite{NSR}, it is given by
\begin{equation}
N=N_0+\frac{1}{\pi}\sum_{\bf q}\int_{-\infty}^\infty d\omega\frac{1}{e^{\beta \omega}-1}\frac{\partial\varphi\left({\bf q},\omega\right)}{\partial\mu}, \label{number}
\end{equation}
where $N_0=2\sum_\mathbf{k}(f^{o}_{\mathbf{k}}+f^{c}_{\mathbf{k}})$ and $\varphi$ is the phase shift in many-body environment given as
\begin{equation}
\varphi({\bf q},\omega) =-\text{Im}\left\{\ln[ 1-g_0(\chi^o+\chi ^c)+(g_0^2-g_1^2)\chi^o\chi^c]\right\}.\label{delta}
\end{equation}
Numerically solving Eq. (\ref{Thouless}) to (\ref{delta}) gives $T_\text{c}$ as a function of $\delta$ for a given density.

Before proceeding to present the numerical results, we should first make two remarks. First, in the limit $\delta=0$, $\chi^o=\chi^c=\chi$, and the Thouless criterion Eq. (\ref{Thouless}) can be reduced to $(1-g_{+}\chi)(1-g_{-}\chi)=0$. In our system with $a_{s+}\gg a_{s-}>0$, it is clear that there will be two solutions. One corresponds to the shallow bound state $-\hbar^2/(ma^2_{s+})$ and the other corresponds to a much deeper bound state $-\hbar^2/(ma^2_{s-})$. These two solutions will also exist for finite $\delta$. However, only the former one is responsible for OFR. In addition, we should note that even if $a_{s-}<0$, as long as $a_{s+}\gg |a_{s-}|$, the OFR phenomenon will still exist. In this case, there is only one bound state and therefore only one solution for $T_\text{c}$. Hence, hereafter we should only focus on the solution that is related to OFR.

Second, as pointed out in Ref. \cite{Zhang}, since there are two scattering channels, one needs to introduce two order parameters in the superfluid phase. Therefore, a natural question is that, starting from the low-temperature superfluid phase and increasing temperature, whether two order parameters will disappear simultaneously or at two different temperatures? By extending the mean-field theory presented in Ref. \cite{Zhang} to finite temperature, one can show that both two order parameters vanish simultaneously. Moreover, similar to the case of the single channel model, by setting both order parameters to zero, the mean-field gap equation is automatically reduced to the Thouless criterion of Eq.~(\ref{Thouless}) \cite{supple}.

\begin{figure}[t]
\includegraphics[width=3.3 in]
{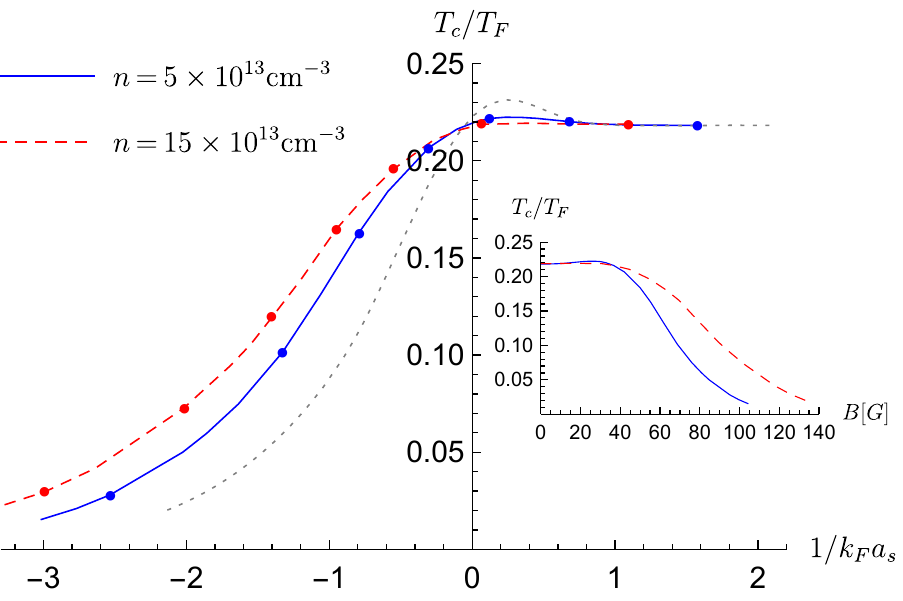}
\caption{The Fermi superfluid transition temperature $T_\text{c}/T_\text{F}$ as a function of $1/(k_\text{F}a_s)$, where $a_s$ is related to $\delta$ via Eq. (\ref{aa}). The blue and red lines and points correspond to two different densities of $5\times 10^{13}\text{cm}^{-3}$ and $15\times 10^{13}\text{cm}^{-3}$, respectively. The solid and dashed lines are NSR calculation based on the full OFR model, while the points are effective single channel calculation with an energy dependent scattering length $a_text{s}(k)$. The dotted line denotes the single-channel
NSR calculation for a wide resonance (with an energy independent $a_\text{s}$). The inset shows $T_\text{c}/T_\text{F}$ as a function of $B$ for $\Delta_m=5$  }
\label{transition}
\end{figure}

Now we present the results for $T_\text{c}$ in Fig.~\ref{transition}. Here $T_\text{c}/T_\text{F}$ is shown as a function of $1/(k_\text{F}a_s)$ for two typical densities. These results are also compared with the results from the single-channel NSR theory with a $k$-independent scattering length $a_{\rm s}$ for a wide resonance. In this single channel calculation, $T_\text{c}/T_\text{F}$ is a function of $1/(k_\text{F}a_s)$ only and is independent of density, while it is clearly not the case for the OFR model. More remarkably, at the BCS side of the resonance, we find that the $T_\text{c}/T_\text{F}$ is higher in the OFR case than in a wide resonance, and the higher the density, the larger $T_\text{c}/T_\text{F}$.

How to understand the higher $T_\text{c}/T_\text{F}$ at the BCS side? Here we show that this is precisely due to the finite range effect of a narrow resonance. To prove this, we perform a NSR calculation using an effective single-channel model but with a $k$-dependent scattering length $a_\text{s}(k)$, that is determined by aforementioned two-body solution of the OFR model \cite{supple}. The results are shown by points in Fig. \ref{transition}. We find that the results coincide perfectly with our NSR calculation with the OFR model. Physically, since $r_0$ is negative, from Eq. (\ref{expansion}) it is clear that as $k$ increases, $-1/a_s(k)$ becomes more negative and effectively the scattering at finite energy is shifted towards the BEC side, corresponding to a stronger attractive interaction. The is also constant with the previous discussion that the narrow resonance has stronger interaction effect based on phase shift \cite{Ho} and higher $T_\text{c}$ using atom-molecule two-channel models \cite{narrowTc1,narrowTc2,narrowTc3,narrowTc4,narrowTc5,narrowTc6}.

\textit{Summary.} We point out that the recently discovered OFR for ${}^{173}$Yb atom is a narrow resonance, because of which the $T_\text{c}$ for the Fermi superfluid at the BCS side is even higher. This prediction of higher $T_\text{c}$, together with the OFR appearing quite wide in term of magnetic field and long lifetime observed in experiments, realizing Fermi superfluid in OFR is quite promising. This new Fermi superfluid will have different many-body properties comparing to that studied in ${}^{40}$K and ${}^6$Li before, such as equation-of-state, collective mode and quasi-particle behaviors. The experimental study of this new Fermi superfluid will be complementary to the previous efforts in the past decade and provide a more comprehensive picture of strongly interacting Fermi gas.

\textit{Acknowledgment.} We thank Boyang Liu, Xiaji Liu, and  Yoji Ohashi for helpful discussions. This work is supported by NSFC Grant No. 11325418 (HZ), No. 11434011(PZ), No. 11504021(JX), and by Tsinghua University Initiative Scientific Research Program (HZ), NKBRSF of China under Grant No. 2012CB922104(PZ), the Research Funds of Renmin University of China under Grant No. 15XNLF18(RQ) and No. 16XNLQ03(PZ), and Fundamental Research Funds for the Central Universities No. FRF-TP-15-040A1(JX).

\begin{newpage}

\begin{appendix}

\begin{widetext}

\section{Appendix}

\subsection{Expansion of Scattering Length around Resonance}

In this section we expand the scattering length $a_{{\rm s}}$ around
the OFR point and prove Eqs. (8-10) of our maintext. The Hamiltonian $\hat{H}=\hat{H}_{0}+\hat{V}$ has been defined in the main text. As point out in Ref.~\cite{Zhang},
the scattering wave function of two atoms incident from the open channel
$|o\rangle$ can be written as
\begin{equation}
|\psi({\bf r})\rangle=\left[e^{i{\bf k}\cdot{\bf r}}+f_{o}(k)\frac{e^{ikr}}{r}\right]|o\rangle+f_{c}(k)\frac{e^{-\sqrt{m\delta/\hbar^{2}-k^{2}}r}}{r}|c\rangle,\label{psir}
\end{equation}
where ${\bf k}$ is the incident relative momentum and $f_{o}(k)$
is the elastic scattering amplitude. Here we consider the cases with $k^2/m<\delta$.
Substituting Eq. (\ref{psir}) into the Schr\"odinger equation $(\hat{H}_{0}+\hat{V})|\psi({\bf r})\rangle=\frac{k^{2}}{m}|\psi({\bf r})\rangle$,
we find that $f_{\text{c}}(k)$ and $f_{\text{o}}(k)$ satisfy equations
\begin{eqnarray}
 & (1+ika_{\text{s}0})f_{o}(k)-a_{\text{s}1}\sqrt{\frac{m\delta}{\hbar^{2}}-k^{2}}f_{\text{c}}(k)+a_{\text{s}0}=0,\label{e1}\\
 & ika_{\text{s}1}f_{\text{o}}(k)+\left(1-a_{\text{s}0}\sqrt{\frac{m\delta}{\hbar^{2}}-k^{2}}\right)f_{\text{c}}(k)+a_{\text{s}1}=0.\label{e2}
\end{eqnarray}
Solving these equation, we obtain the scattering amplitude $f_{\text{o}}(k)$:
\begin{eqnarray}
f_{o}(k)=\frac{-a_{\text{s}0}+(a_{s0}^{2}-a_{s1}^{2})\sqrt{m\delta/\hbar^{2}-k^{2}}}{1+a_{s0}\left(ik-\sqrt{m\delta/\hbar^{2}-k^{2}}\right)-ik(a_{s0}^{2}-a_{s1}^{2})\sqrt{m\delta/\hbar^{2}-k^{2}}}.\label{eqn:nonzerof}
\end{eqnarray}

The zero-energy scattering length $a_{{\rm s}}$ is defined as
\begin{equation}
a_{{\rm s}}=-f_{o}(k=0).\label{sl}
\end{equation}
Thus, the result (\ref{eqn:nonzerof}) yields \cite{Zhang}
\begin{eqnarray}
a_{{\rm s}}=\frac{-a_{s0}+(a_{s0}^{2}-a_{s1}^{2})\sqrt{m\delta/\hbar^{2}}}{a_{s0}\sqrt{m\delta/\hbar^{2}}-1},\label{eqn:zeroas}
\end{eqnarray}
i.e., Eq. (7) of our maintext. It is clear that $a_{{\rm s}}$ is
a function of $\delta$ and diverges when $\delta=\delta_{{\rm res}}\equiv\hbar^{2}/ma_{s0}^{2}$,
i.e., the OFR occurs when $\delta=\delta_{{\rm res}}$. Expanding Eq. \ref{eqn:zeroas} in term of $\delta-\delta_\text{res}$, we obtain
\begin{equation}
a_{{\rm s}}=-\frac{2\hbar^{2}a_{s1}^{2}}{ma_{s0}^{3}}\frac{1}{\left(\delta-\delta_{{\rm res}}\right)}+\frac{\left(2a_{s0}^{2}-3a_{s1}^{2}\right)}{2a_{s0}}+{\cal O}\left(\delta-\delta_{{\rm res}}\right).\label{ass}
\end{equation}
By replacing $\delta=(\delta\mu)B$, Eq. \ref{ass} becomes 
\begin{equation}
a_{{\rm s}}\approx a_{\text{bg}}\left(1-\frac{\Delta_{\text{B}}}{B-B_{\text{res}}}\right)\label{as2}
\end{equation}
where
\begin{align}
 & a_{\text{bg}}=\frac{2a_{s0}^{2}-3a_{s1}^{2}}{2a_{s0}},\label{abg}\\
 & B_{\text{res}}=\frac{\hbar^{2}}{\delta\mu ma_{s0}^{2}},\label{Bres}\\
 & \Delta_{B}=\frac{4\hbar^{2}a_{s1}^{2}}{\delta\mu ma_{s0}^{2}(2a_{s0}^{2}-3a_{s1}^{2})}.\label{DeltaB}
\end{align}
These are Eqs. (8-10) of our maintext.

\subsection{Effective Range Expansion}

Now we calculate the effective range $r_{0}$ and prove Eq. (11) of
our maintext. According to the scattering theory, the scattering amplitude
$f_{o}(k)$ is related to the phase shift $\theta_{k}$ via
\begin{eqnarray}
f_{o}(k)=\frac{-1}{ik-k\cot\theta_{k}}.\label{eqn:general-f}
\end{eqnarray}
Eqs. (\ref{eqn:nonzerof}) and (\ref{eqn:general-f}) give the result
\begin{equation}
-\frac{1}{a_s(k)}\equiv k\cot\theta_{k}=\frac{1+a_{s0}\sqrt{m\delta/\hbar^{2}-k^{2}}}{-a_{s0}+(a_{s0}^{2}-a_{s1}^{2})\sqrt{m\delta/\hbar^{2}-k^{2}}}.\label{kcot}
\end{equation}
Using this result and the relation $k\cot\theta_{k}=-1/a_{{\rm s}}+r_{0}k^{2}/2+{\cal O}(k^{3}),$
i.e., Eq. (1) of our maintext, we obtain the effective range $r_{0}$

\begin{eqnarray}
r_{0}=-\frac{a_{s1}^{2}}{\sqrt{m\delta/\hbar^{2}}\left[a_{s0}-\sqrt{m\delta/\hbar^{2}}\left(a_{s0}^{2}-a_{s1}^{2}\right)\right]^{2}}.
\end{eqnarray}
This is Eq. (11) of our maintext.

\subsection{Bound-State Energy }

Now we study the two-body bound-state energy. There are two bound states in such system, and only the shallow one is
responsible to OFR. Therefore, here we will focus on the shallow bound
state. 

\begin{figure}
\includegraphics[width=4in]{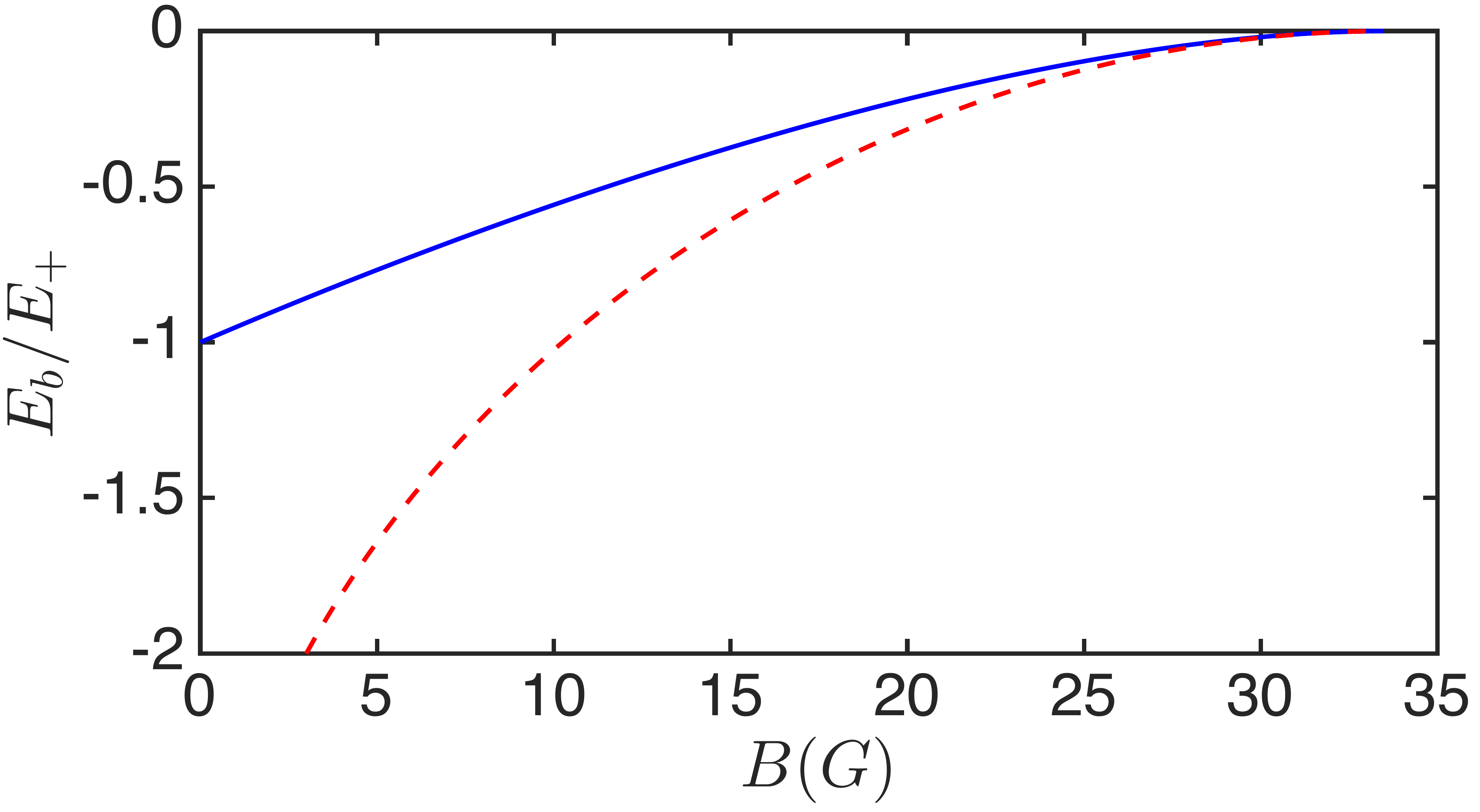}
\caption{
The energy $E_b$ of the shallow two-body bound state,
as a function of magnetic field $B$. Here $E_{+}=\hbar^{2}/(ma_{s+}^{2})$ is introduced as the energy unit.
The blue solid line is the result given by the two-body calculation of the OFR model. The red dashed line
is the reference for a wide resonance.
}

\label{Eb}
\end{figure}


In our system the two-body bound state wave function can be written
as
\begin{equation}
|\psi_{b}({\bf r})\rangle=C_{o}\frac{e^{-\sqrt{-mE_{b}/\hbar^{2}}r}}{r}|o\rangle+C_{c}\frac{e^{-\sqrt{m/\hbar^{2}(\delta-E_{b})}r}}{r}|c\rangle\label{psib}
\end{equation}
where $E_{b}$ is the bound-state energy and $C_{o}$ and $C_{c}$
are constants. Substituting Eq. (\ref{psib}) into the Schr\"odinger
equation $(\hat{H}_{0}+\hat{V})|\psi_{b}({\bf r})\rangle=E_{b}|\psi_{b}({\bf r})\rangle$,
we obtain the equation for the energy $E_{b}$:
\begin{eqnarray}
\det\left(\begin{array}{cc}
1-a_{s0}\sqrt{m(\delta-E_{b})/\hbar^{2}} & -a_{s1}\sqrt{-mE_{b}/\hbar^{2}}\\
-a_{s1}\sqrt{m(\delta-E_{b})/\hbar^{2}} & 1-a_{s0}\sqrt{m(\delta-E_{b})/\hbar^{2}}
\end{array}\right)=0\label{eqn:2c-Eb}
\end{eqnarray}
We can obtain $E_{b}$ by numerically solving Eq.~(\ref{eqn:2c-Eb}).
In Fig.~\ref{Eb} we illustrate $E_{b}$ as a function of $B$. For
comparison, we also present the bound-state energy $-\hbar^{2}/(ma_{{\rm s}}^{2})$
with $a_{{\rm s}}$
given by Eq. (\ref{eqn:zeroas}) which is the binding energy obeyed by a wide resonance.

\subsection{Transition Temperature $T_{c}$}

In this section we present the finite temperature features of mean-field solution for superfluid
nearby OFR. The many-body Hamiltonian of such system can be written
as
\begin{eqnarray}
 & \hat{H}=\sum_{{\bf k}}\left[\varepsilon_{{\bf k}}^{o}(c_{g\downarrow{\bf k}}^{\dagger}c_{g\downarrow{\bf k}}+c_{e\uparrow{\bf k}}^{\dagger}c_{e\uparrow{\bf k}})+\varepsilon_{{\bf k}}^{c}(c_{g\uparrow{\bf k}}^{\dagger}c_{g\uparrow{\bf k}}+c_{e\downarrow{\bf k}}^{\dagger}c_{e\downarrow{\bf k}})\right]+\frac{g_{+}}{2}\hat{A}_{+}^{\dag}\hat{A}_{+}+\frac{g_{-}}{2}\hat{A}_{-}^{\dag}\hat{A}_{-},
\end{eqnarray}
where $\varepsilon_{{\bf k}}^{o}=\hbar^{2}{\bf k}^{2}/(2m)-\mu$,
$\varepsilon_{{\bf k}}^{c}=\hbar^{2}{\bf k}^{2}/(2m)+\delta/2-\mu$,
and
\begin{eqnarray}
 & \hat{A}_{+}=\sum_{{\bf k}}(c_{g\uparrow{\bf -k}}c_{e\downarrow{\bf k}}-c_{g\downarrow{\bf -k}}c_{e\uparrow{\bf k}});\\
 & \hat{A}_{-}=\sum_{{\bf k}}(c_{g\uparrow{\bf -k}}c_{e\downarrow{\bf k}}+c_{g\downarrow{\bf -k}}c_{e\uparrow{\bf k}}).
\end{eqnarray}
By defining
the order parameters $\Delta_{\pm}=g_{\pm}\langle\hat{A}\rangle/2$
and extending the zero-temperature BCS theory to finite temperature,
one obtains the gap and number equations \cite{Zhang}
\begin{eqnarray}
 & -\frac{m}{4\pi\hbar^{2}}\left[\frac{1}{a_{s+}}+\frac{1}{a_{s-}}+\frac{\Delta_{o}}{\Delta_{c}}\left(\frac{1}{a_{s-}}-\frac{1}{a_{s+}}\right)\right]=\sum_{{\bf k}}\left[\frac{1-2f_{{\bf k}}^{c}}{\sqrt{\left(\varepsilon_{{\bf k}}^{c}-\mu\right)^{2}+|\Delta_{c}|^{2}}}-\frac{2m}{\hbar^{2}k^{2}}\right];\label{closedgap}\\
 & -\frac{m}{4\pi\hbar^{2}}\left[\frac{1}{a_{s+}}+\frac{1}{a_{s-}}+\frac{\Delta_{c}}{\Delta_{o}}\left(\frac{1}{a_{s-}}-\frac{1}{a_{s+}}\right)\right]=\sum_{{\bf k}}\left[\frac{1-2f_{{\bf k}}^{o}}{\sqrt{\left(\varepsilon_{{\bf k}}^{o}-\mu\right)^{2}+|\Delta_{o}|^{2}}}-\frac{2m}{\hbar^{2}k^{2}}\right];\label{opengap}\\
 & n=\sum_{{\bf k}}\left[2-\frac{\left(\varepsilon_{{\bf k}}^{c}-\mu\right)(1-2f_{{\bf k}}^{c})}{\sqrt{\left(\varepsilon_{{\bf k}}^{c}-\mu\right)^{2}+|\Delta_{c}|^{2}}}-\frac{\left(\varepsilon_{{\bf k}}^{o}-\mu\right)(1-2f_{{\bf k}}^{o})}{\sqrt{\left(\varepsilon_{{\bf k}}^{o}-\mu\right)^{2}+|\Delta_{o}|^{2}}}\right],\label{num}
\end{eqnarray}
where $\Delta_{c}=\Delta_{-}+\Delta_{+}$ and $\Delta_{o}=\Delta_{-}-\Delta_{+}$
are order parameters in the closed and open channel, respectively. Here
the Fermi distribution function $f_{{\bf k}}^{o/c}$ are define as
\begin{eqnarray}
f_{{\bf k}}^{o/c}=\frac{1}{{\rm e}^{\beta\sqrt{\left(\varepsilon_{{\bf k}}^{o/c}-\mu\right)^{2}+|\Delta_{o/c}|^{2}}}+1}
\end{eqnarray}
By solving Eq.~(\ref{closedgap}), (\ref{opengap}) and Eq.~(\ref{num}),
one can find the paring order parameters decrease monotonously with the increasing of the temperature, as shown in Fig.~\ref{Finiteorder}, and most importantly, $\Delta_{c}$ and $\Delta_{o}$ vanish simultaneously.
\begin{figure}
\includegraphics[width=6in]{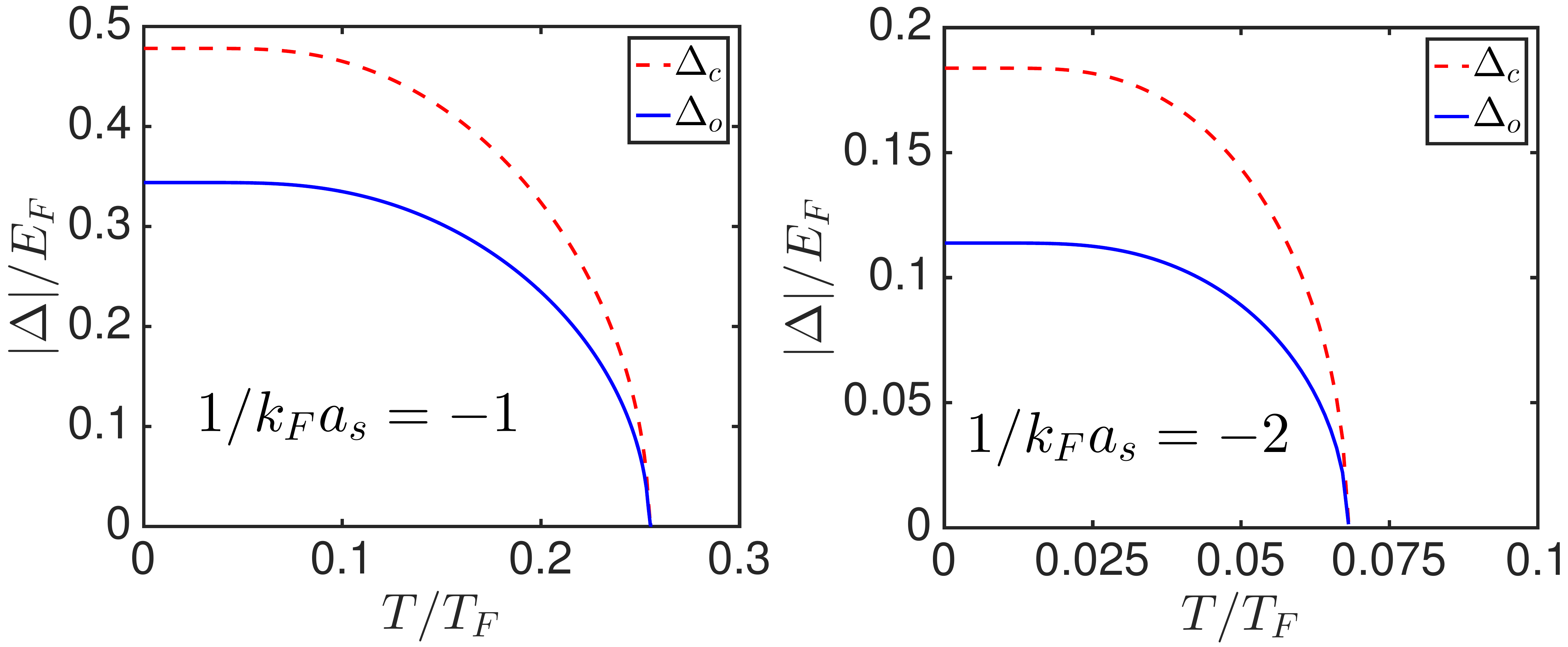} \caption{Paring order parameter $\Delta_{c/o}$ v.s. temperature $T/T_\text{F}$ in BCS
regime. The blue solid line and red dashed line denote the order parameters
in the open and closed channel, respectively. In our calculation, the
number density $n=k_{F}^{3}/(3\pi^{2})$ is taken as $5\times10^{13}$cm$^{-3}$.
The order parameters decrease monotonously with the increasing of temperature
and both two order parameters vanish simultaneously at $T_{c}$.}

\label{Finiteorder}
\end{figure}


Now we prove that by taking $\Delta_{c/o}=0$ in Eq.~(\ref{closedgap})
and (\ref{opengap}), one obtains the Thouless criterion. To this
end, one needs to recast these two equations as
\begin{eqnarray}
 & \frac{\Delta_{o}}{\Delta_{c}}\left(\frac{1}{a_{s-}}-\frac{1}{a_{s+}}\right)=\frac{4\pi\hbar^{2}}{m}\sum_{{\bf k}}\left[\frac{2m}{\hbar^{2}{\bf k}^{2}}-\frac{1-2f_{{\bf k}}^{c}}{\sqrt{\left(\varepsilon_{{\bf k}}^{c}-\mu\right)^{2}+|\Delta_{c}|^{2}}}\right]-\left(\frac{1}{a_{s+}}+\frac{1}{a_{s-}}\right);\label{close}\\
 & \frac{\Delta_{c}}{\Delta_{o}}\left(\frac{1}{a_{s-}}-\frac{1}{a_{s+}}\right)=\frac{4\pi\hbar^{2}}{m}\sum_{{\bf k}}\left[\frac{2m}{\hbar^{2}{\bf k}^{2}}-\frac{1-2f_{{\bf k}}^{o}}{\sqrt{\left(\varepsilon_{{\bf k}}^{o}-\mu\right)^{2}+|\Delta_{o}|^{2}}}\right]-\left(\frac{1}{a_{s+}}+\frac{1}{a_{s-}}\right)\label{open}
\end{eqnarray}
By taking $\Delta_{o/c}=0$, one can define $\chi_{0}^{o/c}$ as
\begin{eqnarray}
\chi_{0}^{{o/c}}=\sum_{{\bf k}}\frac{1-2f_{{\bf k}}^{o/c}}{2\mu-2\varepsilon_{{\bf k}}^{o/c}},
\end{eqnarray}
Here the Fermi distribution functions reduce to $f_{\mathbf{k}}^{o/c}=[\exp(\varepsilon_{{\bf k}}^{o/c})+1]^{-1}$.
Then multiplying Eq.~(\ref{close}) and (\ref{open}), one can obtain
\begin{eqnarray}
1-g_{0}\left(\chi_{0}^{\text{o}}+\chi_{0}^{\text{c}}\right)+\left(g_{0}^{2}-g_{1}^{2}\right)\chi_{0}^{\text{o}}\chi_{0}^{\text{c}}=0,\label{eqn:thouless}
\end{eqnarray}
which is just the Thouless criterion. In the derivation of Eq.~(\ref{eqn:thouless}),
we have used the renormalization $1/g_{\pm}=m/(4\pi\hbar^{2}a_{s\pm})-\sum_{{\bf k}}m/(\hbar^{2}k^{2})$
and the relations $g_{0}=(g_{+}+g_{-})/2$ and $g_{1}=(g_{-}-g_{+})/2$.

\subsection{Effective Single-channel NSR Calculation}

In this section, we investigate the superfluid transition temperature based on an effective single-channel calculation but taking into account the energy dependence of scattering length. Let us start from the Nozi\`eres and Schmitt-Rink (NSR) calculation with a constant scattering length and then show how to include the energy dependence in this formula \cite{Nozieres,Zwerger}.
Considering the following single-channel Hamiltonian
\begin{eqnarray}
H=\sum_{{\bf p}}\varepsilon_{{\bf p}}\left(c_{\downarrow{\bf {p}}}^{\dagger}c_{\downarrow{\bf {p}}}+c_{\uparrow{\bf {p}}}^{\dagger}c_{\uparrow{\bf {p}}}\right)+g\sum_{{\bf {pp'q}}}c_{\uparrow{\bf {q}/2+{\bf p}}}^{\dagger}c_{\downarrow{\bf {q}/2-{\bf {p}}}}^{\dagger}c_{\downarrow{\bf {q}/2-{\bf p'}}}c_{\uparrow{\bf {q}/2+{\bf p'}}},
\end{eqnarray}
where $\varepsilon_{{\bf p}}=\hbar^{2}{\bf {p}}^{2}/(2m)-\mu$,
and $1/g=m/(4\pi\hbar^2 a_s)-\sum_{{\bf p}}m/(\hbar^2 p^2)$ with $a_s$ the (constant)
$s$-wave scattering length, in the NSR approach, the transition temperature $T_c$ is determined by the following Thouless criterion and number equations:
\begin{eqnarray}
&&T^{-1}(0,0)=0,\label{ThoulessNSR}\\
&&N=N_0-\frac{1}{\pi}\sum_{\bf k}\int_{-\infty}^\infty d\omega\frac{1}{e^{\beta \omega}-1}\frac{\partial}{\partial\mu}\text{Im}[\ln T^{-1}({\bf q},\omega)], \label{number}
\end{eqnarray}
where $N_{0}=2\sum_{{\bf p}}f_{{\bf p}}$ with $f_{{\bf p}}=\left[{\rm exp}(\varepsilon_{{\bf p}})+1\right]^{-1}$
being the Fermi distribution and
\begin{eqnarray}
T^{-1}({\bf q},\omega)=\frac{m}{4\pi\hbar^2 a_s}-\chi_R({\bf q},\omega)\label{T-matrix}
\end{eqnarray}
is the inverse of many-body T-matrix and $\chi_R({\bf q},\omega)=\chi({\bf q},\omega)+\sum_{\bf p}m/(\hbar^2p^2)$ is the renormalized particle-particle bubble with:
\begin{eqnarray}
\chi({\bf q},\omega)=\sum_{\bf p}\chi_{\bf p}({\bf q},\omega)=\sum_{{\bf p}}\frac{1-f_{{\bf {q}/2+{\bf {p}}}}-f_{{\bf {q}/2-{\bf {p}}}}}{\omega+i0^+-\varepsilon_{{\bf {q}/2+{\bf {p}}}}-\varepsilon_{{\bf {q}/2-{\bf {p}}}}}.
\end{eqnarray}

Now, to include the energy dependent scattering length, we first write the fully interaction strength $g_{\bf pp'}({\bf q},\omega)$ and the $T$-matrix is written diagrammatically as
\begin{eqnarray}
T_{\bf pp'}({\bf q},\omega)=g_{\bf pp'}({\bf q},\omega)+\sum_{\bf p''}g_{\bf pp''}({\bf q},\omega)\chi_{\bf p''}({\bf q},\omega)T_{\bf p''p'}({\bf q},\omega).\label{dia-T}
\end{eqnarray}
Considering that the interaction strength is only dependent of relative energy of colliding particles, we can write $g_{\bf pp'}({\bf q},\omega)=g({\bf q},\omega)=g(E)$, where $E=\omega+2\mu-\hbar^2q^2/(4m)$ is the relative colliding energy due to energy conservation. In this case Eq. (\ref{dia-T}) gives
\begin{align}
T^{-1}_{\bf pp'}({\bf q},\omega)=T^{-1}({\bf q},\omega)=1/g(E)-\chi({\bf q},\omega).
\end{align}
Using the renormalization relation $1/g(E=\hbar^2k^2/m)=m/\left[4\pi\hbar^2a_s(k)\right]-\sum_{\bf p}m/(\hbar^2p^2)$ we finally include the energy dependent scattering in the $T$-matrix as
\begin{eqnarray}
T^{-1}({\bf q},\omega)=\frac{m}{4\pi\hbar^2 a_s(\sqrt{(\omega+2\mu)m/\hbar^{2}-q^{2}/4})}-\chi_R({\bf q},\omega),\label{T-matrix2}
\end{eqnarray}
where the function $a_s(k)$ is given in Eq. (\ref{kcot}). Inserting Eq. (\ref{T-matrix2}) into (\ref{ThoulessNSR}) and (\ref{number}) completes our single channel NSR formula with energy dependent scattering length. The corresponding numerical results for $T_c$ are shown in Fig. 3 (points) of the maintext.

\end{widetext}

\end{appendix}

\end{newpage}

\end{document}